\documentclass{Interspeech2024}

\usepackage{bbm}
\usepackage{amssymb,amsmath}
\usepackage{xcolor}
\usepackage{booktabs} 
\usepackage{algorithm}
\usepackage[noend]{algpseudocode}
\usepackage{url}
\usepackage{graphicx}
\usepackage{caption}
\usepackage{subcaption}
\usepackage{adjustbox}

\newcommand{\x}{\mathbf{x}}

\newcommand{\p}{\mathbf{p}}

\renewcommand{\t}{\mathbf{t}}
\renewcommand{\u}{\mathbf{u}}

\newcommand{\X}{\mathcal{X}}
\newcommand{\V}{\mathcal{V}}
\newcommand{\K}{\mathcal{K}}
\newcommand{\R}{\mathbb{R}}
\newcommand{\ourmethod}{KG-Whisper}
\newcommand{\ourmethodb}{KG-Whisper-PT}

\interspeechcameraready

\title{Keyword-Guided Adaptation of Automatic Speech Recognition} 

\name[]{Aviv}{Shamsian}
\name[]{Aviv}{Navon}
\name[]{Neta}{Glazer}
\name[]{Gill}{Hetz}
\name[]{Joseph}{Keshet}

\address{
  $ $aiOla Research, Israel}
\email{\{aviv.shamsian, aviv\}@aiola.com}


\begin{document}

\maketitle

\begin{abstract}

Automatic Speech Recognition (ASR) technology has made significant progress in recent years, providing accurate transcription across various domains. However, some challenges remain, especially in noisy environments and specialized jargon. In this paper, we propose a novel approach for improved jargon word recognition by contextual biasing Whisper-based models. We employ a keyword spotting model that leverages the Whisper encoder representation to dynamically generate prompts for guiding the decoder during the transcription process. We introduce two approaches to effectively steer the decoder towards these prompts: KG-Whisper, which is aimed at fine-tuning the Whisper decoder, and  KG-Whisper-PT, which learns a prompt prefix. Our results show a significant improvement in the recognition accuracy of specified keywords and in reducing the overall word error rates. Specifically, in unseen language generalization, we demonstrate an average WER improvement of $5.1\%$ over Whisper.


\end{abstract}

\section{Introduction}



Automatic speech recognition (ASR) capabilities have rapidly progressed in recent years through advancements in deep learning models trained on massive datasets of 
speech. Models like HuBERT~\cite{hsu2021hubert} and wav2vec~\cite{baevski2020wav2vec} utilize self-supervised pre-training techniques to learn representations of acoustic and linguistic patterns from hundreds of thousands of hours of diverse audio data. 
Although such models have shown great potential in producing rich representations useful for transcription, their lack of a high-quality decoder limits their robustness and generalization. Recently, \cite{radford2023robust} alleviated this limitation by presenting a new encoder-decoder transformer architecture named Whisper. Whisper was trained on a massive multilingual corpus of transcribed web audio data comprising 680,000 hours of speech. 

However, Whisper and other state-of-the-art ASR models may suffer from performance degradation when applied to real-world datasets presenting numerous challenges~\cite{gong2023whisper}.
For instance, in industrial surroundings where heavy machinery operates continuously, background noise can reach high levels. This acoustic environment makes it difficult for ASR systems to accurately transcribe spoken commands or safety alerts issued by workers. Similarly, on public transportation like a crowded bus or subway train, ambient noise from engines or conversations can interfere with passenger attempts to a voice-activated information system. Additionally, domain-specific utterances and terminology, pose unique challenges for ASR systems, as they require the identification of specialized vocabulary and language patterns. For instance, in the medical domain, physicians detailing medical conditions or treatment plans, and in legal contexts, where lawyers and judges employ legal jargon, case citations, and procedural language uncommon in everyday conversation, demonstrate such challenges.
To develop speech recognition systems that are useful in real-world situations, it is essential to overcome these practical hurdles.

One approach for alleviating these challenges is utilizing domain-specific or personalized context for biasing transcriptions using prior knowledge. A common approach is injecting context into beam search or using shallow fusion~\cite{williams2018contextual,zhao2019shallow,gourav2021personalization,eitan2023combining}, deep fusion~\cite{pundak2018deep,alon2019contextual,chang2021context,sathyendra2022contextual,sainath2023improving}, or a combination of deep and shallow biasing~\cite{le2021deep,le2021contextualized,xu2023cb,sun2023can}. Recently, \cite{liao2023zero} proposed adapting Whisper into specific domains using domain prompts which shows substantial reductions in word error rates. However, this method requires finetuning with a large amount of data and is limited in terms of the prior knowledge and terminology one can provide. Prompting and prompt tuning Whisper was also shown beneficial for adapting to novel tasks, such as audio-visual
speech recognition and code switching~\cite{peng2023prompting}, and target speaker ASR~\cite{ma2023extending}. However, the challenge remains to design an efficient and effective way for contextually biasing Whisper using domain-specific and personalized terminology.




\begin{figure*}[t]
\centering
\begin{subfigure}{.5\textwidth}
  \centering
  \includegraphics[height=7cm]{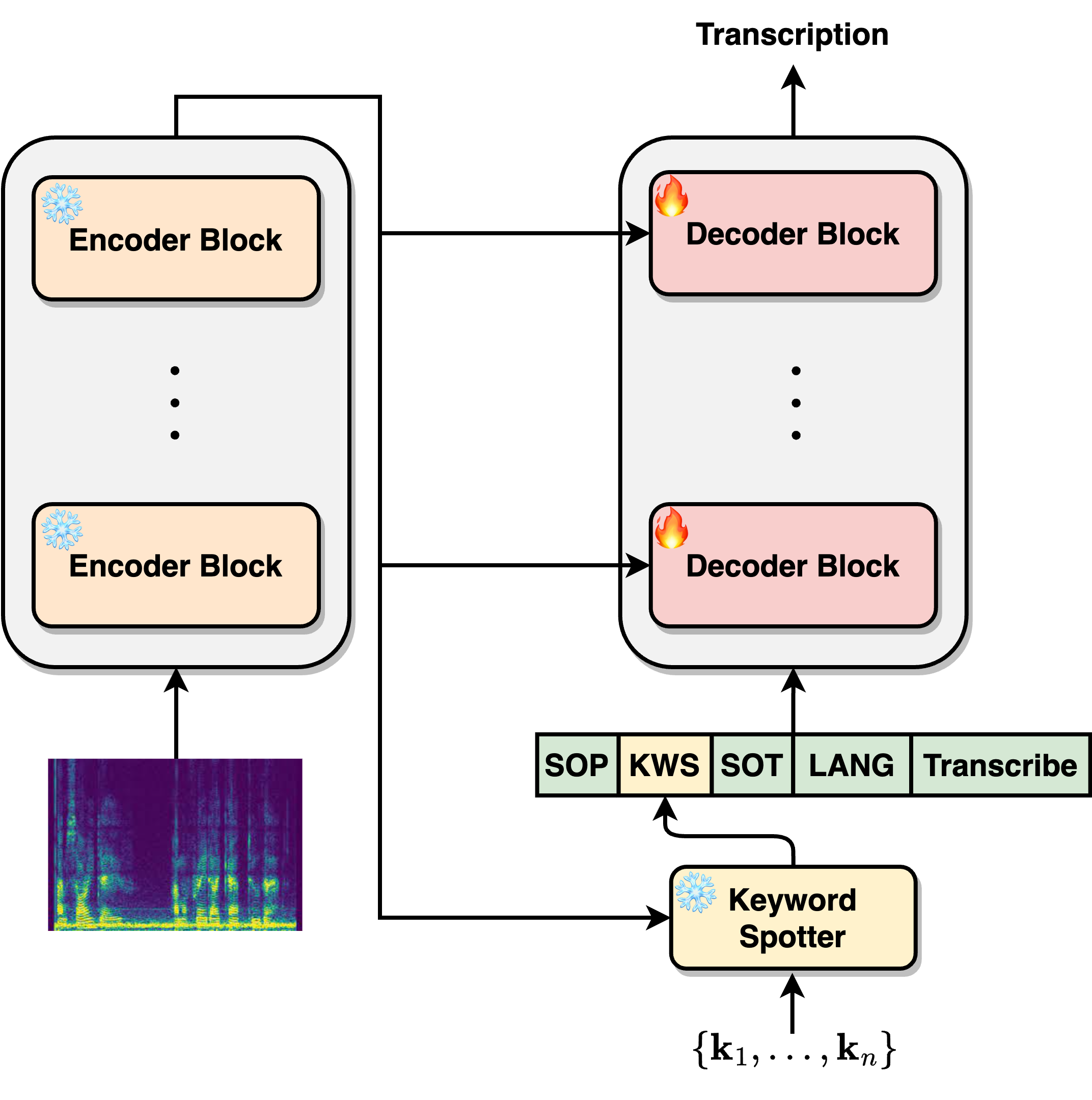}
  \caption{\ourmethod{} - decoder finetuning.}
  \label{fig:kg-whisper}
\end{subfigure}%
\begin{subfigure}{.5\textwidth}
  \centering
  \includegraphics[height=7cm]{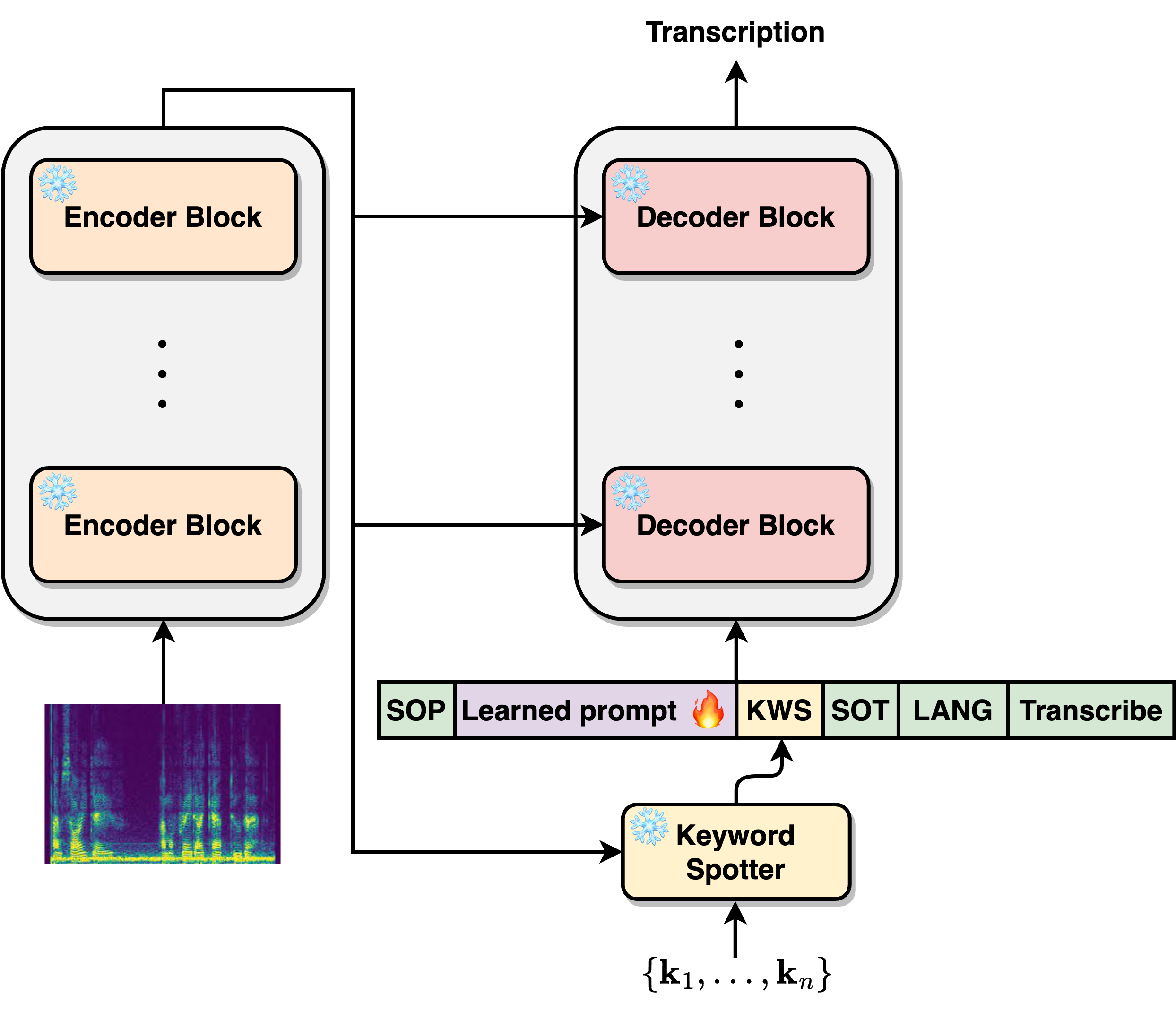}
  \caption{\ourmethodb{} - prompt tuning. }
  \label{fig:kg-whisper-pt}
\end{subfigure}
\caption{An illustration of (a) \ourmethod{} - the decoder receives keywords during fine-tuning, while the encoder module remains frozen. (b) \ourmethodb{} - the entire Whisper model parameters are frozen and only a small number of prompt tokens are tuned.}
\label{fig:arch}
\end{figure*}

To address this issue, we propose a novel approach for contextual biasing. We utilize recent advancements in open vocabulary keyword spotting~\cite{shin2022learning,nishu2023matching,nishu2023flexible} to guide the ASR predictions using prior domain knowledge as a source of contextual information. Our approach first identifies domain-specific and personalized jargon using AdaKWS~\cite{navon2024open}, a KWS system built on top of Whisper's acoustic encoder. Next, the identified keywords are used to prompt the decoder to encourage their incorporation into the transcribed text. We propose two approaches for finetune Whisper for using keyword prompts. The first, termed \ourmethod{} for keyword-guided Whisper, finetunes the entire set of decoder parameters. The second and more efficient approach based on prompt tuning~\cite{lester2021power}, termed \ourmethodb{}, achieves on-par or superior generalization results using only ${\sim} 15$K trainable parameters. 
This approach preserves the original pre-trained Whisper parameters, mitigating potential degradation resulting from fine-tuning over specific datasets and enhancing the model's ability to generalize to novel datasets and domains. 
Most similar to our approach is the concurrent work~\cite{Li2023AMT}. However, the paper provides only 
small-scale finetune experiments on a limited number of languages, or employ spoken form prompts.


Throughout extensive evaluation, we empirically demonstrate that \ourmethod{} and \ourmethodb{} constantly outperform natural baselines. Furthermore, we evaluate the generalization of the proposed method to novel domains and challenging acoustic settings. Finally, we evaluate the generalization performance to novel languages not seen during training. 




\section{Method}

Our goal is to reduce the overall Word Error Rate (WER) of the final prediction and increase the recall over domain-specific (jargon) terms. To achieve this, we integrate the predictions of a Keyword Spotting (KWS) model into an ASR system, guiding the ASR predictions towards the set of detected keywords. 

We first present the notations and learning setup. Let $\X^T$ denote the domain of sequences of $T$ speech frames. Let $\V$ denote a dictionary of tokens (sub-word units), and $\V^L$ the domain of all token sequences containing $L$ tokens (sub-word units). 
Each token is represented by a $D$-dimensional vector.
 
We utilize Whisper \cite{radford2023robust}, a state-of-the-art transformer-based ASR. Its encoder converts a speech sequence into a meaningful sequential representation, taking advantage of the self-attention mechanism. Given a speech input sequence $\x \in \X^T$, consisting of $T$ speech frames, the encoder function $\u = f^e_{\phi}(\x)$ produces an output $\u \in \R^F$. Here, $F$ represents the dimension of the acoustic representation, and $\phi$ denotes the encoder parameters.
 
The transformer decoder function takes the encoder representation $\u$, a prompt $\p \in \V^L$, and the previous tokens, $\t^{i-1}$. It then predicts the next token $\hat{t}^i = f^d_{\psi}(\mathbf{u}, \mathbf{p}, \t^{i-1})$, where $\hat{t}^i\in\V$ and $\psi$ represents the set of decoder parameters. Given a training set of $M$ examples, $\mathcal{S}=\{(\x_j, \p_j, \t_j)\}_{j=1}^{M}$, the ASR is trained to minimize the empirical loss:
\begin{equation}
    \min_{\{\phi, \psi\}} \sum_j \sum_i \mathcal{L}^{\text{CE}}\big( 
    f^d_{\psi}(f^e_{\phi}(\x_j), \p_j, \t^{i-1}_{j} ), t^i_j
    \big)~, 
\end{equation}
where $\mathcal{L}^{\text{CE}}(\hat{t}, t)$ is the cross-entropy loss between the a predicted token, $\hat{t}$, and the correct one, $t$.

Expanding upon Whisper's architecture and optimization, our goal is to guide its predictions toward specific keywords (which can represent out-of-vocabulary or rare words), while preserving the overall performance of the rest of the words in the language. Our approach is based on utilizing decisions from an open vocabulary KWS model~\cite{navon2024open}. 

\begin{figure*}[t]
     \centering
     \begin{subfigure}[b]{0.44\textwidth}
         \centering
         \includegraphics[width=0.9\textwidth]{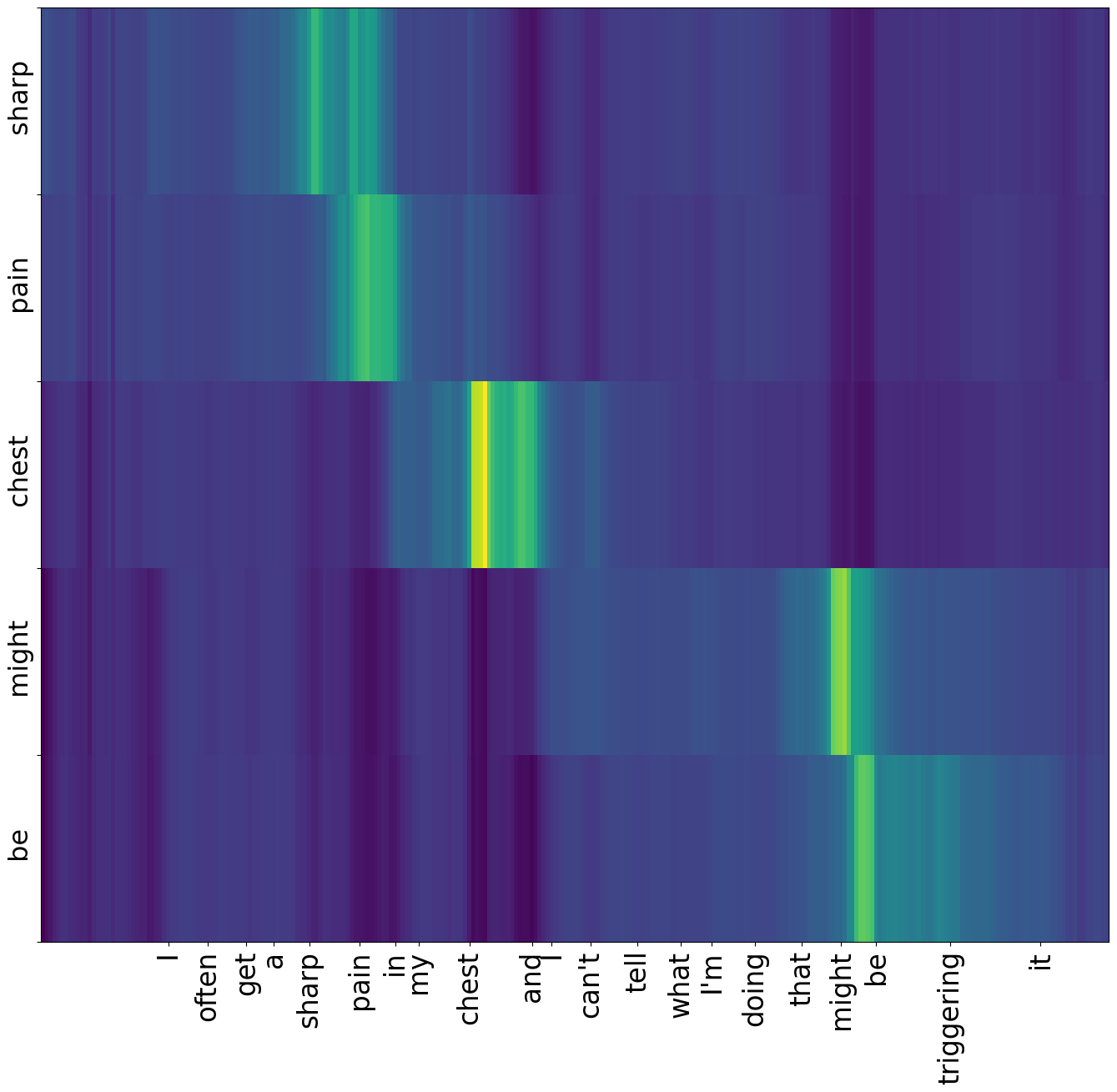}
         \caption{KG-Whisper}
         \label{fig:attn-kg-whisper-ft}
     \end{subfigure}
     \hfill
     \begin{subfigure}[b]{0.44\textwidth}
         \centering
    \includegraphics[width=0.9\textwidth]{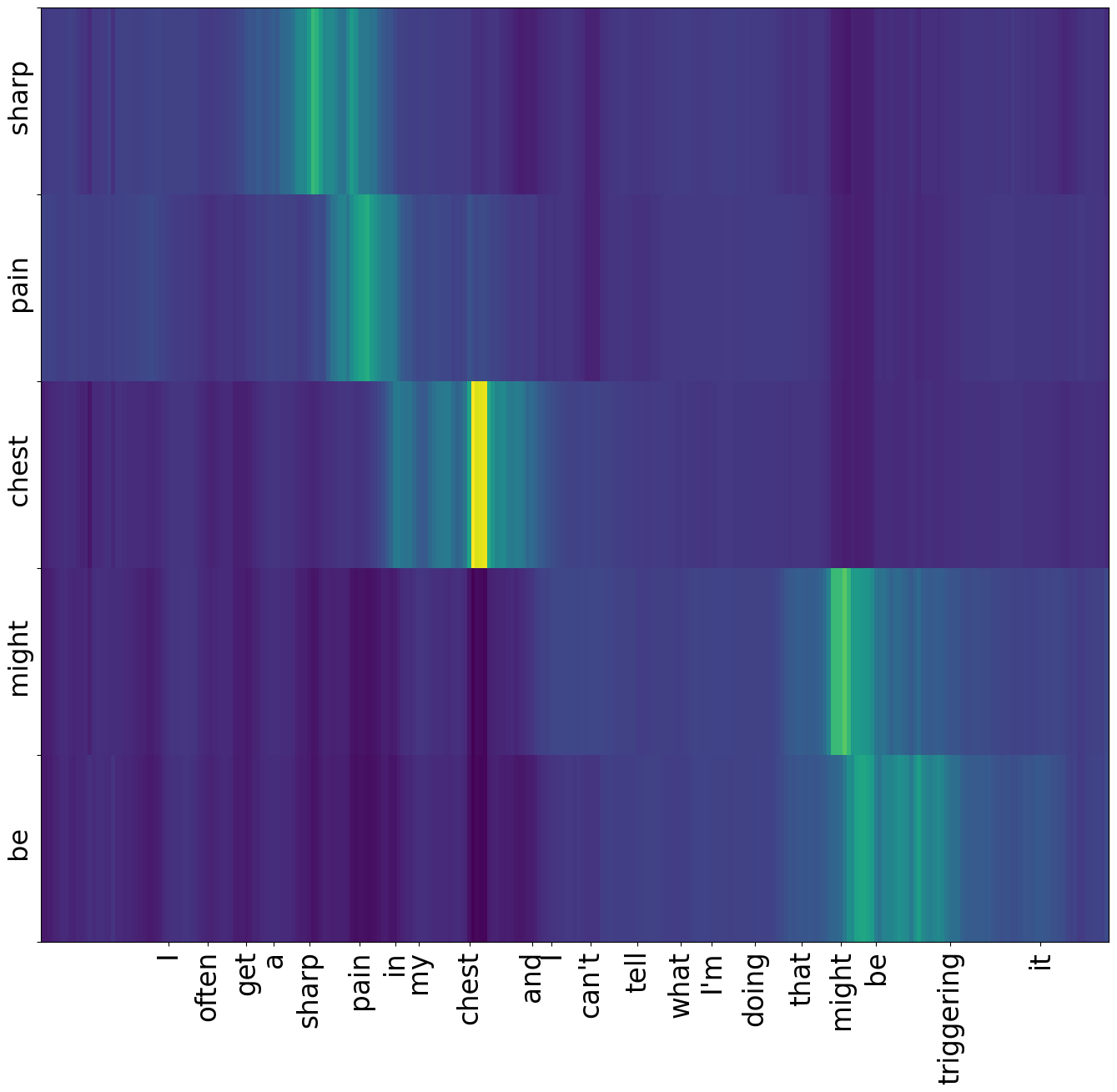}
         \caption{KG-Whisper-PT}
         \label{fig:attn-kg-whisper-pt}
     \end{subfigure}
 \vspace{-5pt}
\caption{Visualization of the cross-attention weights for \ourmethod{} and \ourmethodb{}. We illustrate how the model's attention is directed towards the identified keywords (Y-axis) as it predicts them within the transcribed text (X-axis).}
\end{figure*}

Let $\K$ represent a set of keywords, with each keyword $k \in \K$ defined as a sequence of tokens within the token vocabulary $\mathcal{V}^L$. The KWS \cite{navon2024open} employs a function $\hat{\mathbf{y}} = f^k_{\theta}(\mathbf{u}, \K)$ that takes as input the encoder representation $\mathbf{u}$ and the set of keywords $\K$, and predicts a binary vector $\hat{\mathbf{y}}\in\{0,1\}^{|\K|}$. This vector indicates the presence or absence of each keyword within the input utterance. The KWS parameters, $\theta$, are optimized to minimize the cross-entropy loss: 
\begin{equation}
    \min_{\theta} ~ \sum_j \sum_{(k_j^+,k_j^-) \in \K } \mathcal{L}^{\text{CE}}\big( 
    f^k_{\theta}(\u_j, \K )
    \big) ~.
\end{equation}

The KWS prediction vector $\hat{\mathbf{y}}$ is then used to generate a prompt that is passed to the Whisper decoder. We introduce two novel approaches for guiding Whisper's predictions toward the identified keywords. In the first approach, \emph{KG-Whisper}, we fine-tune Whisper's decoder while freezing the encoder. In the second approach, \emph{KG-Whisper-PT}, we tune the prompt by adding a learned prefix to it, while freezing the encoder the decoder. In both approaches we keep the KWS model freezed. 

\subsection{KG-Whisper: Keyword-guided tuned-decoder}

Our first approach involves fine-tuning the decoder parameters $\psi$. Here, each training example is a tuple of a speech representation $\u$, a prompt $\p^{\text{KWS}(\u, \K)}$ generated from the KWS, the previous (correct) tokens $\t^{i-1}$, and the next token $t^i$:
\begin{equation}
    \min_{\psi} ~ \sum_j \sum_i \mathcal{L}^{\text{CE}}\big( 
    f^d_{\psi}(\u_j, \p^{\text{KWS}(\u_j, \K)}, \t^{i-1}_{j} ), t^i_j
    \big)~.
\end{equation}
Namely, we fine-tune the decoder parameters to minimize the CE loss over pairs of speech representations and KWS prompts.

\subsection{KG-Whisper-PT: Keyword-guided prompt-tuning}

\begin{table}
\centering
\caption{WER and F1 results for Voxpopuli test dataset.}
\begin{tabular}{lccc}
    \toprule
 &\multicolumn{2}{c}{Voxpopuli}\\
        \cmidrule{2-4}
         &  WER $\downarrow$ & F1 $\uparrow$ & \# Params\\
         \midrule
         \ourmethod{} -- Oracle & $9.41$ & $97.53$ & $906$M\\
         \ourmethodb{} -- Oracle & $10.86$ & $95.91$ & $15$K\\
         \midrule
         Whisper &  $13.39$ & $81.77$ & $-$\\
         Whisper + prompt&  $16.70$ & $83.77$ & $-$\\
         Whisper FT &  $11.57$ & $82.54$ & $906$M\\
         Whisper PT & $12.83$ & $90.38$ & $15$K\\
         \midrule
         \ourmethod &  $\mathbf{9.78}$& $91.54$ & $906$M\\
         \ourmethod{}-PT &  $11.10$ & $\mathbf{95.25}$ & $15$K \\
         \bottomrule
    \end{tabular}
    \label{tab:vox}
\end{table}

In the second method, we learn a prefix to the prompt that the KWS generates, $\mathbf{q}\in\R^{N\times D}$, composed of $N$ tokens, where each token is a vector in $\R^D$. This is done by minimizing the CE loss, while keeping the Whisper's encoder and decoder frozen. 
\begin{equation}
    \min_{\mathbf{q}} ~ \sum_j \sum_i \mathcal{L}^{\text{CE}}\big( 
    f^d_{\psi}(\u_j, [\mathbf{q} ~ \p^{\text{KWS}(\u_j, \K)}], \t^{i-1}_{j} ), t^i_j
    \big)~,
\end{equation}
where $\psi$ denote the frozen parameters of the decoder.

In this paper, we highlight the application of contextual biasing with a keyword spotter, focusing specifically on Whisper. However, we note that both approaches outlined in this work apply to any encoder-decoder-based ASR system.

\subsection{Training strategy}

Throughout the training phase, we emulate the KWS predictions by dynamically generating a collection of positive and negative keywords for each audio input within the batch. First, we randomly select the quantity of keywords in the prompt, varying between $1$ to $5$. Next, for each keyword, we determine if the keyword is positive or negative through a coin flip, with a $0.9$ probability assigned to positive keywords. Finally, we determine the keyword length by randomly sampling the number of tokens for each keyword term, ranging from $1$ to $4$. Positive keywords are sampled from the transcript of the input audio sample, while negative keywords are generated from other randomly selected samples within the batch. After sampling the keywords we concatenate them with $\vert$ delimiter. To pass the keyword as context to the decoder module, we place them between the $\langle \textit{start\_of\_prev} \rangle \textit{(SOP)}$ and $\langle \textit{start\_of\_transcript} \rangle \textit{(SOT)}$ tokens, as depicted in Figure~\ref{fig:arch}.

\section{Experiments}
In this section, we evaluate \ourmethod{} and \ourmethodb{} using several datasets and learning setups. 
\\\textbf{Datasets.} 
We use several well-known datasets. \textit{Voxpopuli}~\cite{wang2021voxpopuli}: a comprehensive collection of recordings from the European Parliament, making it a diverse speech dataset.
The dataset includes more than $1600$ hours of annotated speech from $16$ languages. \textit{UWB-ATCC}~\cite{zuluaga2023does}: The Ultra-Wideband Air Traffic Control Communications dataset contains radio communications between aircraft pilots and air traffic control systems. This dataset is extremely challenging for ASR models due to high levels of noise exceeding $20 \text{dB}$. \textit{Medical}~\cite{kumar2022deep}: collection of medical reports made by patients including over $6600$ utterances. 
\textit{Fleurs}~\cite{conneau2023fleurs}: a diverse multilingual dataset with $12$ hours of annotated audio from $102$ languages.

\noindent\textbf{Data preprocessing.} We follow the data preprocessing paradigm presented by~\cite{radford2023robust, navon2024open}. We resample the audio samples to $16$Khz followed by log-magnitude Mel-spectrogram generation. Specifically, we generate 80-channel Mel-spectrograms using 25-millisecond windows and 10-millisecond stride.

\noindent\textbf{Experimental setup.} We train \ourmethod{} and \ourmethodb{} on Voxpopuli dataset for $30$K update steps using batch size of $4$ and $1e{-}7$ and $5e{-}4$ learning rate, respectively. Unless stated otherwize, we use a 12 token prefix prompt for \ourmethodb{}. We use a frozen version (i.e., without trainable parameters) of AdaKWS~\cite{navon2024open} as the keyword spotter.
In the evaluation step, we query the keyword spotter with $20$ keywords for each audio sample. Among these keywords, $3$ are positive, meaning they appear in the audio utterance, while the remaining $17$ are negative. The keywords are sampled proportionally to the term frequency–inverse document frequency~\cite{salton1988term} scores. 

We report common metrics in ASR domain, namely Word Error Rate (WER), and the F1 score for the positive keywords. 

\noindent\textbf{Baselines.} We compare our methods to several natural baselines. The compared methods include: (1) \textit{Whisper} - a pre-trained Whisper large-v2 model. (2) \textit{Whisper + prompt} - providing keywords detected by the keyword spotter as prompt to a pre-trained Whisper large-v2 model. (3) \textit{Whisper FT} - Whisper large-v2 model with fine-tuned decoder. (4) \textit{\ourmethod{}-Oracle} - \ourmethod{} with optimal KWS, i.e., providing the positive ground truth keywords as prompt. (5) \textit{\ourmethodb{}-Oracle} - \ourmethodb{} with optimal KWS.

\begin{table}[t]
\small
\centering
\caption{\textit{Out of domain generalization}: WER and F1 results for UWB-ATCC and Medical test datasets.}
\begin{adjustbox}{max width=0.48\textwidth}
\begin{tabular}{lccccc}
\toprule
 & \multicolumn{2}{c}{UWB-ATCC} &  & \multicolumn{2}{c}{Medical} \\ \cmidrule{2-3} \cmidrule{5-6} 
 & WER $\downarrow$ & F1 $\uparrow$  && WER $\downarrow$  & F1 $\uparrow$    \\
 \midrule
 Whisper&      $76.55$       &    $24.54$       &  &  $7.33$             &   $80.50$          \\
 Whisper + prompt&      $\mathbf{55.27}$       &    $61.57$       &  & $32.82$              &  $68.71$ \\
 \midrule
 \ourmethod &   $56.09$ & $67.75$ && $6.87$ &  $88.74$ \\
 \ourmethod{}-PT &  $57.88$ & $\mathbf{70.89}$  &  & $\mathbf{6.15}$  & $\mathbf{96.58}$ \\
 \bottomrule
\end{tabular}
\end{adjustbox}
\label{tab:atc_med}
\end{table}


\subsection{Multi-lingual Finetuning} First, we evaluate \ourmethod{} and \ourmethodb{} using the Voxpopuli dataset. We use the training set to fine-tune the decoder of \ourmethod{}, or the prefix prompt of \ourmethodb{}, and evaluate the performance using the test set. The results are presented in Table~\ref{tab:vox}. \ourmethod{} surpasses Whisper variations by a notable margin, demonstrating enhancements of $1.8\%$ in WER and $9\%$ in F1 metrics compared to the top-performing baseline. The increase in F1 score suggests that our method effectively guides Whisper towards the provided context keywords, resulting in a higher frequency of positive keyword appearances in the predicted transcript compared to other baselines. Furthermore, \ourmethodb{} shows a $2.3\%$ reduction in Whisper's WER, while utilizing a mere $15$K learnable parameters. 
Moreover, both \ourmethod{} and \ourmethodb{} achieve comparable results to their Oracle baseline counterparts which applied to the ground truth keywords as prompt. Remarkably, Whisper + prompt exhibits inferior performance compared to the naive Whisper. We observed that this discrepancy stems from the fact that presenting prompts without fine-tuning encourages Whisper to generate hallucinated content.


\subsection{Out of domain generalization}
We further investigate \ourmethod{} and \ourmethodb{} ability to generalize to out-of-domain datasets characterized by challenging acoustic environments, domain-specific jargon words, and new languages. Here, we use the \ourmethod{} and \ourmethodb{} models trained on the Voxpopuli dataset. Notably, our evaluation involves datasets \emph{not used} during the training process of the ASR and the KWS model, namely zero-shot learning setup. 
First, we evaluate our methods on UWB-ATCC~\cite{zuluaga2023does} dataset which presents acoustic challenges due to poor signal-to-noise ratio. The noise, originating from radio communication, reaches over $30$dBs at peak, making UWB-ATCC challenging even for state-of-the-art ASR systems like Whisper. Next, using the Medical~\cite{kumar2022deep} dataset, we test how well \ourmethod{} and \ourmethodb{} handle out-of-domain jargon words. Specifically, the Medical dataset contains jargon words from the medical domain which we assume that Whisper was less exposed to during training. Finally, we randomly select $6$ low-resource languages from the Fleurs~\cite{conneau2023fleurs} dataset to evaluate generalization to unseen languages that were not represented in Whisper's training data (Estonian: ET, Hungarian: HU , Latvian: LV, Slovenian: SI, and Uzbek: UZ). The results are presented in Tables \ref{tab:atc_med} and \ref{tab:fleurs}. We show that \ourmethodb{} achieves substantial improvement over Whisper baselines, improving both WER and keywords F1 on UWB-ATCC and Medical datasets. Additionally, our method enhances Whisper's WER performance on all randomly selected languages from the Fleurs dataset. Overall, \ourmethodb{} yields an average WER improvement of $5.1\%$ over Whisper.



\begin{table}[t]
\centering
\caption{\textit{Generalization to unseen languages}: WER results for unseen languages from Fleurs dataset.}
\begin{adjustbox}{max width=0.48\textwidth}
\begin{tabular}{lcccccccc}
\toprule
                 & ET & HU & LV & SI & UZ \\
\midrule
Whisper\hspace{-6px}          &  $24.96$      &     $19.14$      &   $25.91$     &        $25.82$           &   $86.69$    \\
Whisper + prompt\hspace{-6px} &    $21.18$      &  $23.19$         &    $23.94$      &   $27.95$        &    $81.30$   \\
\midrule
\ourmethod{}\hspace{-6px}  &    $18.39$  &   $15.90$  & $22.44$  &    $21.24$        &  $\mathbf{80.39}$  \\  
\ourmethod{}-PT\hspace{-6px} & $\mathbf{17.67}$ & $\mathbf{15.68}$ &  $\mathbf{21.95}$ & $\mathbf{20.98}$ & $80.94$ \\
\bottomrule
\end{tabular}
\end{adjustbox}
\label{tab:fleurs}
\end{table}


\begin{table}[h]
\centering
\caption{Learned prompt ablation: test results on Voxpopuli and Medical datasets with varying number of learned tokens.}
\begin{tabular}{lccccc}
    \toprule
 &\multicolumn{2}{c}{Voxpopuli} &&\multicolumn{2}{c}{Medical} \\
        \cmidrule{2-3} \cmidrule{5-6}
          & WER $\downarrow$ & F1 $\uparrow$  && WER $\downarrow$ & F1 $\uparrow$ \\
         \midrule
         4 tokens & $11.72$ & $94.66$ && $6.42$ & $96.37$ \\
         8 tokens & $11.54$ & $94.97$ && $6.69$ & $96.22$ \\
         12 tokens &  $\mathbf{11.10}$ & $95.25$ && $\mathbf{6.15}$ & $96.58$ \\
         16 tokens&  $11.29$ & $\mathbf{95.35}$ && $6.17$ & $\mathbf{96.61}$\\
         20 tokens &  $11.27$ & $95.18$ && $6.27$ & $96.38$ \\
         24 tokens &  $11.95$ &$94.11$ && $6.91$ & $95.62$\\
         \bottomrule
    \end{tabular}
    \label{tab:vox_ablation}
\end{table}

\subsection{Ablation study}
We empirically investigate the effect of learned prompt length on the performance of \ourmethodb{}. Specifically, we evaluate \ourmethodb{} on Voxpopuli and Medical datasets while varying the number of learnable tokens ranging from $4$ to $24$ tokens. The results, presented in Table \ref{tab:vox_ablation}, 
demonstrates that the WER increases when the prompt length either decreases or increases, with the best WER and F1 results are shown for 12 and 16 tokens. For longer prompts, the decoder tends to introduce more widely insertion errors. These errors originated in \emph{hallucinations}~\cite{koenecke2024careless}, whereas undesirable text is generated. In addition, long prompts increase the computational demand of the decoder. In contrast, shorter prompts provide limited context for the decoder, causing it to omit words or part of the original speech, which leads to an increase in deletion errors.

\section{Conclusion}
Whisper is almost as good as a human listener for recognizing speech under varying environmental conditions. However, its performance declines in specialized spoken domains characterized by rare words or jargon language. This paper introduces two novel techniques designed to enhance Whisper's detection of specific words by leveraging keyword spotting. Utilizing prior knowledge on the domain and keyword spotter, our method provides context to Whisper and boosts its transcription capabilities. Both methods outperform Whisper baselines on various datasets, even in challenging acoustic environments. Additionally, we demonstrate that our methods can generalize to novel languages. We are confident that our paper can lead to further research aimed at improving the resilience and robustness of speech foundation models.  

\bibliographystyle{IEEEtran}
\bibliography{mybib}

\begin{thebibliography}{10}
\providecommand{\url}[1]{#1}
\csname url@samestyle\endcsname
\providecommand{\newblock}{\relax}
\providecommand{\bibinfo}[2]{#2}
\providecommand{\BIBentrySTDinterwordspacing}{\spaceskip=0pt\relax}
\providecommand{\BIBentryALTinterwordstretchfactor}{4}
\providecommand{\BIBentryALTinterwordspacing}{\spaceskip=\fontdimen2\font plus
\BIBentryALTinterwordstretchfactor\fontdimen3\font minus \fontdimen4\font\relax}
\providecommand{\BIBforeignlanguage}[2]{{%
\expandafter\ifx\csname l@#1\endcsname\relax
\typeout{** WARNING: IEEEtran.bst: No hyphenation pattern has been}%
\typeout{** loaded for the language `#1'. Using the pattern for}%
\typeout{** the default language instead.}%
\else
\language=\csname l@#1\endcsname
\fi
#2}}
\providecommand{\BIBdecl}{\relax}
\BIBdecl

\bibitem{hsu2021hubert}
W.-N. Hsu, B.~Bolte, Y.-H.~H. Tsai, K.~Lakhotia, R.~Salakhutdinov, and A.~Mohamed, ``Hubert: Self-supervised speech representation learning by masked prediction of hidden units,'' \emph{IEEE/ACM Transactions on Audio, Speech, and Language Processing}, vol.~29, pp. 3451--3460, 2021.

\bibitem{baevski2020wav2vec}
A.~Baevski, Y.~Zhou, A.~Mohamed, and M.~Auli, ``wav2vec 2.0: A framework for self-supervised learning of speech representations,'' \emph{Advances in neural information processing systems}, vol.~33, pp. 12\,449--12\,460, 2020.

\bibitem{radford2023robust}
A.~Radford, J.~W. Kim, T.~Xu, G.~Brockman, C.~McLeavey, and I.~Sutskever, ``Robust speech recognition via large-scale weak supervision,'' in \emph{International Conference on Machine Learning (ICML)}.\hskip 1em plus 0.5em minus 0.4em\relax PMLR, 2023, pp. 28\,492--28\,518.

\bibitem{gong2023whisper}
Y.~Gong, S.~Khurana, L.~Karlinsky, and J.~Glass, ``Whisper-at: Noise-robust automatic speech recognizers are also strong general audio event taggers,'' \emph{arXiv preprint arXiv:2307.03183}, 2023.

\bibitem{williams2018contextual}
I.~Williams, A.~Kannan, P.~S. Aleksic, D.~Rybach, and T.~N. Sainath, ``Contextual speech recognition in end-to-end neural network systems using beam search.'' in \emph{Interspeech}, 2018, pp. 2227--2231.

\bibitem{zhao2019shallow}
D.~Zhao, T.~N. Sainath, D.~Rybach, P.~Rondon, D.~Bhatia, B.~Li, and R.~Pang, ``Shallow-fusion end-to-end contextual biasing.'' in \emph{Interspeech}, 2019, pp. 1418--1422.

\bibitem{gourav2021personalization}
A.~Gourav, L.~Liu, A.~Gandhe, Y.~Gu, G.~Lan, X.~Huang, S.~Kalmane, G.~Tiwari, D.~Filimonov, A.~Rastrow \emph{et~al.}, ``Personalization strategies for end-to-end speech recognition systems,'' in \emph{ICASSP 2021-2021 IEEE International Conference on Acoustics, Speech and Signal Processing (ICASSP)}, 2021, pp. 7348--7352.

\bibitem{eitan2023combining}
D.~Eitan, M.~Pirchi, N.~Glazer, S.~Meital, G.~Ayach, G.~Krendel, A.~Shamsian, A.~Navon, G.~Hetz, and J.~Keshet, ``Combining language models for specialized domains: A colorful approach,'' \emph{arXiv preprint arXiv:2310.19708}, 2023.

\bibitem{pundak2018deep}
G.~Pundak, T.~N. Sainath, R.~Prabhavalkar, A.~Kannan, and D.~Zhao, ``Deep context: end-to-end contextual speech recognition,'' in \emph{2018 IEEE spoken language technology workshop (SLT)}, 2018, pp. 418--425.

\bibitem{alon2019contextual}
U.~Alon, G.~Pundak, and T.~N. Sainath, ``Contextual speech recognition with difficult negative training examples,'' in \emph{ICASSP 2019-2019 IEEE International Conference on Acoustics, Speech and Signal Processing (ICASSP)}, 2019, pp. 6440--6444.

\bibitem{chang2021context}
F.-J. Chang, J.~Liu, M.~Radfar, A.~Mouchtaris, M.~Omologo, A.~Rastrow, and S.~Kunzmann, ``Context-aware transformer transducer for speech recognition,'' in \emph{2021 IEEE Automatic Speech Recognition and Understanding Workshop (ASRU)}, 2021, pp. 503--510.

\bibitem{sathyendra2022contextual}
K.~M. Sathyendra, T.~Muniyappa, F.-J. Chang, J.~Liu, J.~Su, G.~P. Strimel, A.~Mouchtaris, and S.~Kunzmann, ``Contextual adapters for personalized speech recognition in neural transducers,'' in \emph{ICASSP 2022-2022 IEEE International Conference on Acoustics, Speech and Signal Processing (ICASSP)}, 2022, pp. 8537--8541.

\bibitem{sainath2023improving}
T.~N. Sainath, R.~Prabhavalkar, D.~Caseiro, P.~Rondon, and C.~Allauzen, ``Improving contextual biasing with text injection,'' in \emph{ICASSP 2023-2023 IEEE International Conference on Acoustics, Speech and Signal Processing (ICASSP)}, 2023, pp. 1--5.

\bibitem{le2021deep}
D.~Le, G.~Keren, J.~Chan, J.~Mahadeokar, C.~Fuegen, and M.~L. Seltzer, ``Deep shallow fusion for rnn-t personalization,'' in \emph{2021 IEEE Spoken Language Technology Workshop (SLT)}, 2021, pp. 251--257.

\bibitem{le2021contextualized}
D.~Le, M.~Jain, G.~Keren, S.~Kim, Y.~Shi, J.~Mahadeokar, J.~Chan, Y.~Shangguan, C.~Fuegen, O.~Kalinli \emph{et~al.}, ``Contextualized streaming end-to-end speech recognition with trie-based deep biasing and shallow fusion,'' \emph{arXiv preprint arXiv:2104.02194}, 2021.

\bibitem{xu2023cb}
Y.~Xu, B.~Liu, Q.~Huang, X.~Song, Z.~Wu, S.~Kang, and H.~Meng, ``Cb-conformer: Contextual biasing conformer for biased word recognition,'' in \emph{ICASSP 2023-2023 IEEE International Conference on Acoustics, Speech and Signal Processing (ICASSP)}, 2023, pp. 1--5.

\bibitem{sun2023can}
G.~Sun, X.~Zheng, C.~Zhang, and P.~C. Woodland, ``Can contextual biasing remain effective with whisper and gpt-2?'' \emph{arXiv preprint arXiv:2306.01942}, 2023.

\bibitem{liao2023zero}
F.-T. Liao, Y.-C. Chan, Y.-C. Chen, C.-J. Hsu, and D.-s. Shiu, ``Zero-shot domain-sensitive speech recognition with prompt-conditioning fine-tuning,'' in \emph{2023 IEEE Automatic Speech Recognition and Understanding Workshop (ASRU)}, 2023, pp. 1--8.

\bibitem{peng2023prompting}
P.~Peng, B.~Yan, S.~Watanabe, and D.~Harwath, ``Prompting the hidden talent of web-scale speech models for zero-shot task generalization,'' \emph{arXiv preprint arXiv:2305.11095}, 2023.

\bibitem{ma2023extending}
H.~Ma, Z.~Peng, M.~Shao, J.~Li, and J.~Liu, ``Extending whisper with prompt tuning to target-speaker asr,'' \emph{arXiv preprint arXiv:2312.08079}, 2023.

\bibitem{shin2022learning}
H.-K. Shin, H.~Han, D.~Kim, S.-W. Chung, and H.-G. Kang, ``Learning audio-text agreement for open-vocabulary keyword spotting,'' \emph{arXiv preprint arXiv:2206.15400}, 2022.

\bibitem{nishu2023matching}
K.~Nishu, M.~Cho, and D.~Naik, ``Matching latent encoding for audio-text based keyword spotting,'' \emph{arXiv preprint arXiv:2306.05245}, 2023.

\bibitem{nishu2023flexible}
K.~Nishu, M.~Cho, P.~Dixon, and D.~Naik, ``Flexible keyword spotting based on homogeneous audio-text embedding,'' \emph{arXiv preprint arXiv:2308.06472}, 2023.

\bibitem{navon2024open}
A.~Navon, A.~Shamsian, N.~Glazer, G.~Hetz, and J.~Keshet, ``Open-vocabulary keyword-spotting with adaptive instance normalization,'' in \emph{Proceeding of the International Conference on Audio, Speech, and Signal Processing (ICASSP)}, 2024.

\bibitem{lester2021power}
B.~Lester, R.~Al-Rfou, and N.~Constant, ``The power of scale for parameter-efficient prompt tuning,'' \emph{arXiv preprint arXiv:2104.08691}, 2021.

\bibitem{Li2023AMT}
\BIBentryALTinterwordspacing
Y.~Li, Y.~Li, M.~Zhang, C.~Su, M.~Ren, X.~Qiao, X.~Zhao, M.~Piao, J.~Yu, X.~Lv, M.~Ma, Y.~Zhao, and H.~Yang, ``A multitask training approach to enhance whisper with contextual biasing and open-vocabulary keyword spotting,'' 2023. [Online]. Available: \url{https://api.semanticscholar.org/CorpusID:266999141}
\BIBentrySTDinterwordspacing

\bibitem{wang2021voxpopuli}
C.~Wang, M.~Riviere, A.~Lee, A.~Wu, C.~Talnikar, D.~Haziza, M.~Williamson, J.~Pino, and E.~Dupoux, ``Voxpopuli: A large-scale multilingual speech corpus for representation learning, semi-supervised learning and interpretation,'' \emph{arXiv preprint arXiv:2101.00390}, 2021.

\bibitem{zuluaga2023does}
J.~Zuluaga-Gomez, A.~Prasad, I.~Nigmatulina, S.~S. Sarfjoo, P.~Motlicek, M.~Kleinert, H.~Helmke, O.~Ohneiser, and Q.~Zhan, ``How does pre-trained wav2vec 2.0 perform on domain-shifted asr? an extensive benchmark on air traffic control communications,'' in \emph{2022 IEEE Spoken Language Technology Workshop (SLT)}, 2023, pp. 205--212.

\bibitem{kumar2022deep}
Y.~Kumar, A.~Koul, and S.~Mahajan, ``A deep learning approaches and fastai text classification to predict 25 medical diseases from medical speech utterances, transcription and intent,'' \emph{Soft computing}, vol.~26, no.~17, pp. 8253--8272, 2022.

\bibitem{conneau2023fleurs}
A.~Conneau, M.~Ma, S.~Khanuja, Y.~Zhang, V.~Axelrod, S.~Dalmia, J.~Riesa, C.~Rivera, and A.~Bapna, ``Fleurs: Few-shot learning evaluation of universal representations of speech,'' in \emph{2022 IEEE Spoken Language Technology Workshop (SLT)}, 2023, pp. 798--805.

\bibitem{salton1988term}
G.~Salton and C.~Buckley, ``Term-weighting approaches in automatic text retrieval,'' \emph{Information processing \& management}, vol.~24, no.~5, pp. 513--523, 1988.

\bibitem{koenecke2024careless}
A.~Koenecke, A.~S.~G. Choi, K.~Mei, H.~Schellmann, and M.~Sloane, ``Careless whisper: Speech-to-text hallucination harms,'' \emph{arXiv preprint arXiv:2402.08021}, 2024.

\end{thebibliography}

\end{document}